\documentclass[conference]{IEEEtran}
\IEEEoverridecommandlockouts
\usepackage{cite}
\usepackage{amsmath,amssymb,amsfonts}
\usepackage{algorithm}
\usepackage{algpseudocode}
\usepackage{graphicx}
\usepackage{textcomp}
\usepackage{xcolor}
\usepackage{subcaption}
\usepackage{url}
\def\BibTeX{{\rm B\kern-.05em{\sc i\kern-.025em b}\kern-.08em
    T\kern-.1667em\lower.7ex\hbox{E}\kern-.125emX}}
\makeatletter
\def\@ConfName{Conferences Name}
\newcommand{\ConfName}[1]{\gdef\@ConfName{\small\sffamily#1}}

\def\@ConfAcronym{Conferences Acronym}
\newcommand{\ConfAcronym}[1]{\gdef\@ConfAcronym{\small\sffamily#1}}

\def\@ConfDate{Conferences Date}
\newcommand{\ConfDate}[1]{\gdef\@ConfDate{\small\sffamily#1}}

\def\@ConfCity{Conferences City}
\newcommand{\ConfCity}[1]{\gdef\@ConfCity{\small\sffamily#1}}

\def\@PaperNo{Paper Number}
\newcommand{\PaperNo}[1]{\gdef\@PaperNo{\large\sffamily#1}}

\def\@SetAffiliation{}
\newcommand{\SetAffiliation}[1]{\gdef\@SetAffiliation{\small#1}}

\def\@SetAuthors{}
\newcommand{\SetAuthors}[1]{\gdef\@SetAuthors{#1}}

\def\@Sponser{}
\newcommand{\Sponser}[1]{\gdef\@Sponser{#1}}

\def\@maketitle{%
	\newpage
	\null
	\vskip 2em%
	\begin{flushright}
		{\bfseries\sffamily \@ConfName\\[3pt]
		\@ConfAcronym\\
		\@ConfDate,
		\@ConfCity}\\
		\vspace*{1.0cm}
	{\bfseries\sffamily\Large	\@PaperNo}\\
	\vspace*{0.5cm}
	\end{flushright}
	\begin{center}%
		\let \footnote \thanks
		{\bfseries\LARGE\sffamily \@title \par}%
		\vskip 1.5em%
		{
			\lineskip .5em%
			\begin{tabular}[t]{c}%
				{\@SetAuthors}
			\end{tabular}\par}
		\vskip 0.5em%
	\@SetAffiliation
		\vskip 1em%
		{\large \@date}%
		\thanks{\noindent\rule{0.3\textwidth}{0.5pt}}
		\thanks{\@Sponser}
	\end{center}%
	\par
	\vskip 1.5em}

\newcommand{\norm}[1]{\left\lVert#1\right\rVert}

\makeatother
\renewcommand\IEEEkeywordsname{Keywords-component} 
\usepackage{footnote}
\renewcommand{\figurename}{Figure.}
\renewcommand{\tablename}{TABLE.}
\begin{document}
\title{TOWARDS A GPU-NATIVE ADAPTIVE MESH REFINEMENT SCHEME FOR THE LATTICE BOLTZMANN METHOD IN COMPLEX GEOMETRIES}

\SetAuthors{
	Khodr Jaber$^1*$,
	Ebenezer E. Essel$^2$,
	Pierre E. Sullivan$^{1,1}$,
	}

\SetAffiliation{
	$^1$Department of Mechanical and Industrial Engineering, University of Toronto, Toronto, Canada\\
	$^2$Department of Mechanical, Industrial and Aerospace Engineering, Concordia University, Montreal, Canada\\
	$^*$khodr.jaber@mail.utoronto.ca
	}

\ConfName{Proceedings of the Canadian Society for Mechanical Engineering International Congress\\
          32nd Annual Conference of the Computational Fluid Dynamics Society of Canada\\
          Canadian Society of Rheology Symposium}
\ConfAcronym{CSME/CFDSC/CSR 2025}
\ConfDate{May 25--28, 2025} 
\ConfCity{Montréal, Québec, Canada} 
\PaperNo{N/A}


\date{February 17, 2025}

\maketitle

\begin{abstract}
\normalfont


We present a GPU-native mesh adaptation procedure that incorporates a complex geometry represented with a triangle mesh within a primary Cartesian computational grid organized as a forest of octrees. A C++/CUDA program implements the procedure for execution on a single GPU as part of a new module with the AGAL framework, which was originally developed for GPU-native adaptive mesh refinement (AMR) and fluid flow simulation with the Lattice Boltzmann Method (LBM). Traditional LBM is limited to grids with regular prismatic cells with domain boundaries aligned with the cell faces. This work is a first step towards an implementation of the LBM that can simulate flow over irregular surfaces while retaining both adaptation of the mesh and the temporal integration routines entirely on the GPU. Geometries can be inputted as a text file (which generates primitive objects such as circles and spheres) or as an STL file (which can be generated by most 3D modeling software). The procedure is divided into three steps: 1) an import step where the geometry is loaded into either an index list arrangement or directly as a face-vertex coordinates list, 2) a spatial binning step where the faces are distributed to a set of bins with user-defined density, and 3) a near-wall refinement step where the cells of the computational grid detect adjacency to the faces stored in the appropriate bin to form the links between the geometry and the boundary nodes. We validate the implementation and assess its performance in terms of total execution time and speedup relative to a serial CPU implementation using a 2D circle and a 3D Stanford bunny.

\end{abstract}

\vspace*{0.5em}

\section{INTRODUCTION} 

Adaptive mesh refinement (AMR) provides a flexible non-uniformity to the underlying grid in computational physics simulations that can be adjusted at runtime. This is especially important for computational fluid dynamics simulations based on the Lattice Boltzmann Method (LBM) --- an emerging competitor to conventional numerical methods for weakly-compressible flow that is well-known for its suitability for parallel implementation with GPUs --- which is traditionally limited to uniform grids due to the discretization procedure. The LBM has been applied to phase-field modeling \cite{Sakane2023}, free-surface flow \cite{Watanabe2021}, and external flows \cite{Suss2023}. Many grid refinement schemes have been developed in the past two decades for the LBM; we refer the reader to the works of Gendre et al. \cite{Gendre2017} and Schukmann et al. \cite{Schukmann2023} for further reading. Several open-source packages implement the LBM with grid refinement such as Palabos \cite{Palabos}, waLBerla \cite{waLBerla}, and ESPResSo \cite{ESPResSo}.

\begin{figure*}[t]
    \centering
    \begin{subfigure}[t]{0.31\textwidth}
        \centering
        \includegraphics[width=1\linewidth]{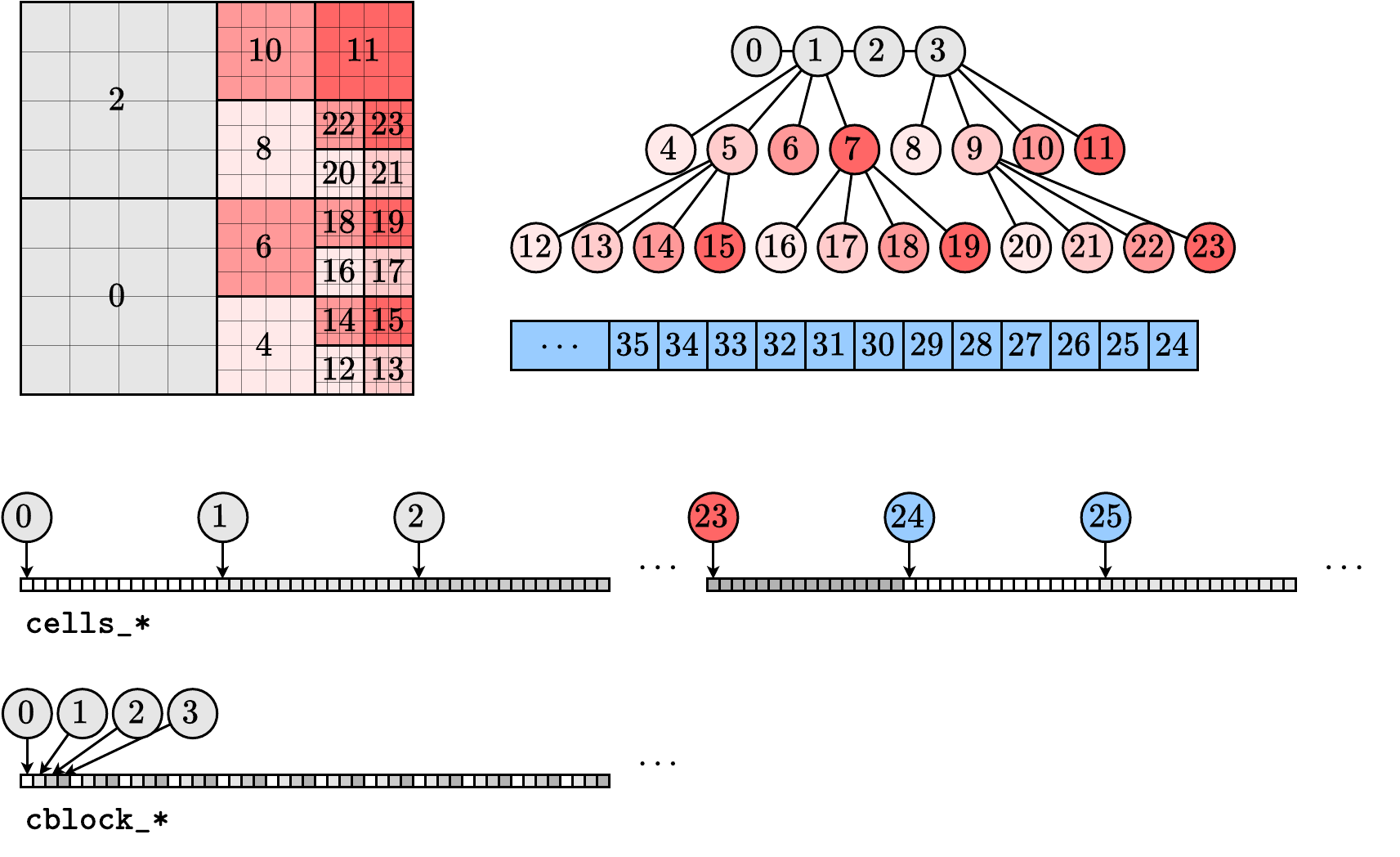}
        \caption{Forest-of-octrees grid in 2D.}
        \label{fig:foo}
    \end{subfigure}
    \hspace{2.5cm}
    \begin{subfigure}[t]{0.31\textwidth}
        \centering
        \includegraphics[width=1\linewidth]{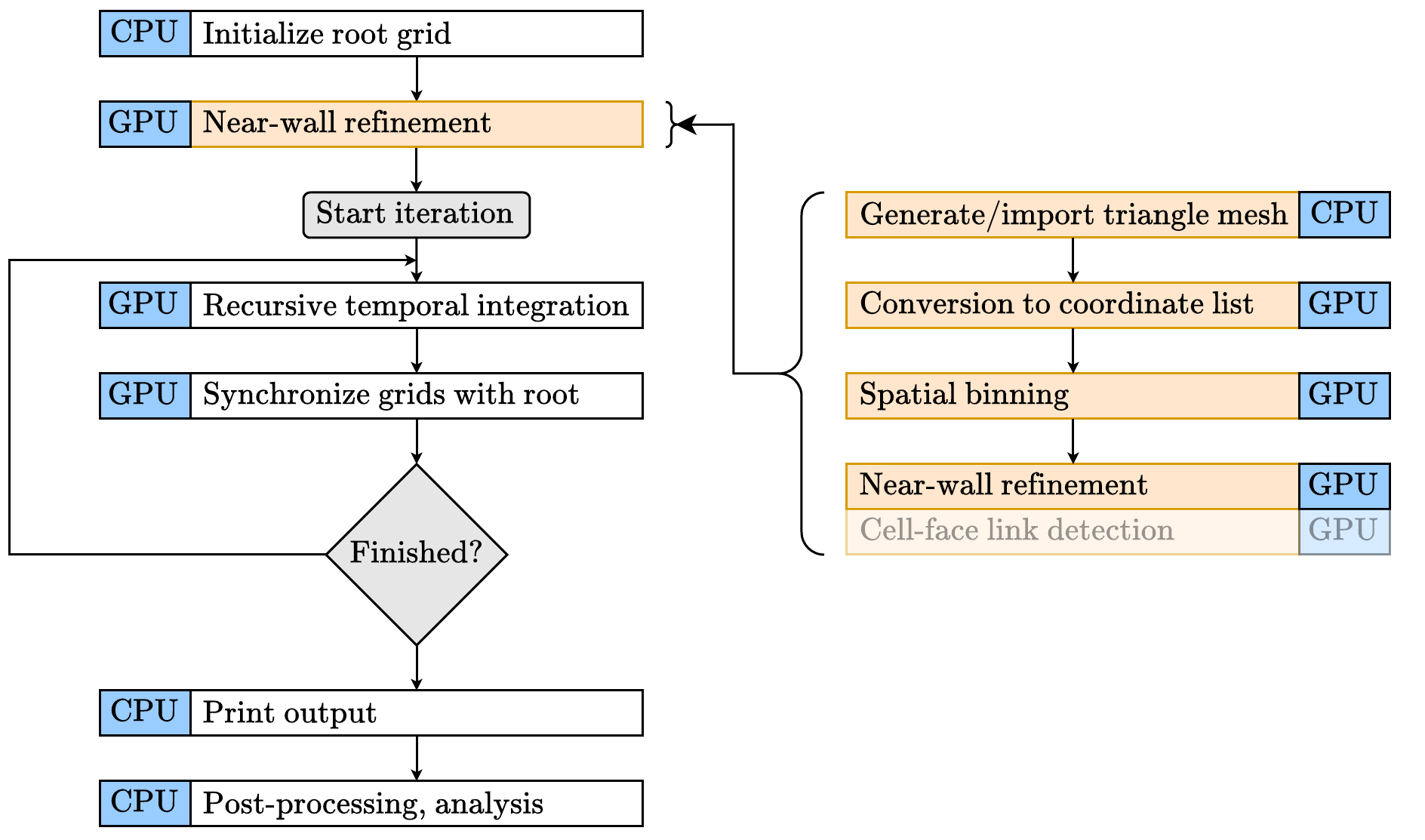}
        \caption{Interaction of \texttt{Mesh} and \texttt{Geometry} modules.}
        \label{fig:scope}
    \end{subfigure}
    \caption{Left: Nodes in the tree correspond to blocks in the grid and identify locations in the data arrays. Right: scope of the current work. The \texttt{Geometry} module replaces the old near-wall refinement routine and enables grid cells to identify the proper domain boundary.}
    \label{fig:current}
\end{figure*}

An AMR scheme in computational fluid dynamics consists of the mesh adaptation routines, which refine the grid according to a user-specified criterion based on the underlying geometry or physics, and the solver routines, which describe coarse-fine grid coupling along refinement interfaces during temporal integration. GPU-acceleration of AMR has also become popular due to advancements in hardware and the increased availability of heterogeneous high-performance computing platforms offered by packages such as AMReX \cite{AMReX} and Daino \cite{Daino}. It is common for the mesh to be managed and adapted on the CPU, and the data retained and advanced entirely on the GPU due to the challenges associated with organizing the required structures on the GPU. However, GPU-native AMR has recently been achieved, where the mesh and data are both processed on the GPU with specialized algorithms and data structures \cite{Giuliani2019,Wang2024,Pavlukhin2024}. In the context of the LBM, the available software packages offering dynamic adaptation of the mesh with GPU-acceleration continue to use the hybrid approach where the mesh is hosted on the CPU, to the best of our knowledge.

We recently developed our own GPU-native AMR approach tailored for Lattice Boltzmann simulations \cite{Jaber2024}, however, the code was limited to square/cubic domains. This work presents a first step towards the integration of complex geometries (in the form of edge/triangle meshes) within our AMR framework for improved and more realistic fluid flow simulations with the LBM.

\section{METHODOLOGY}

This section describes the data structures and algorithms that represent the complex geometry and its incorporation in the forest-of-octree computational grid.

    \begin{figure}[b]
        \centering
        \includegraphics[width=0.6\linewidth,trim={0 0 3cm 3cm}]{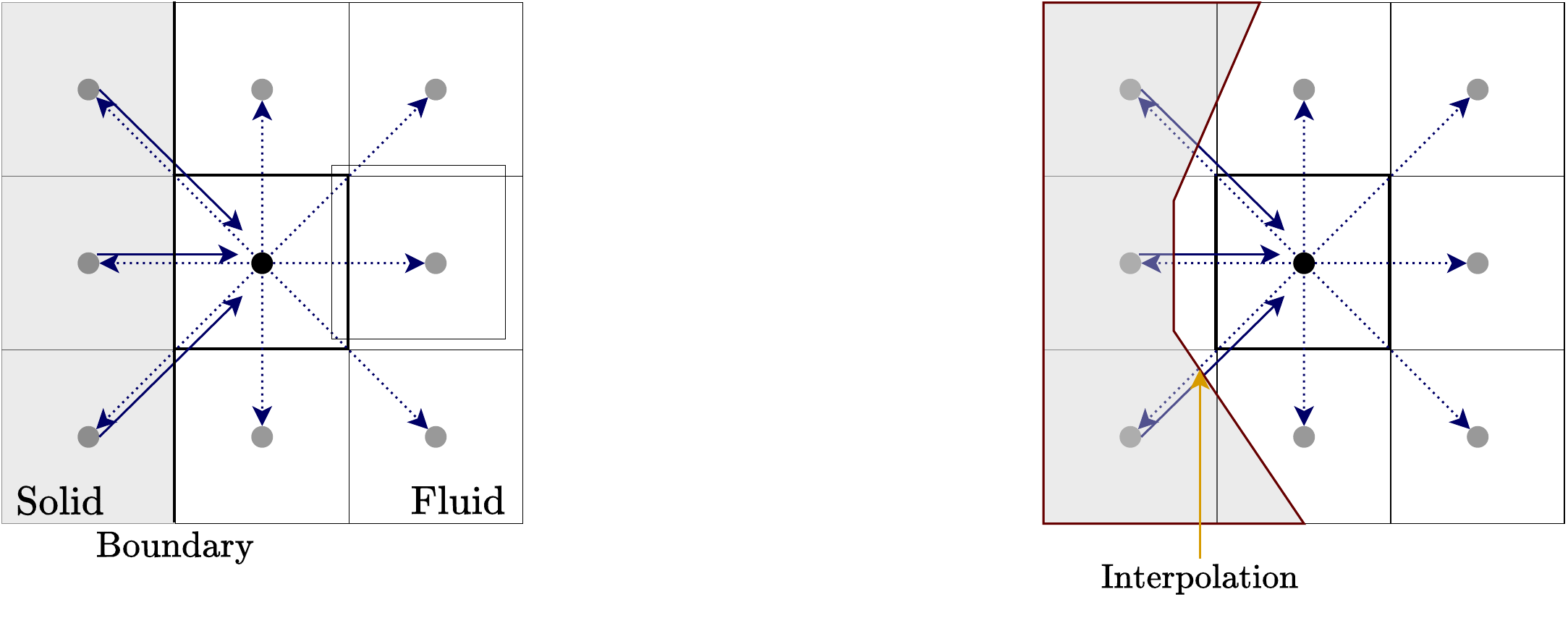}
        \caption{Half-way bounce-back boundary conditions are second-order accurate for geometries aligned with the cell edges/faces, but interpolation or ghost methods are needed to maintain accuracy for more general surfaces.}
        \label{fig:lbm_interp}
    \end{figure}

    \subsection{Adaptive Mesh Refinement on GPUs}

    In an earlier work \cite{Jaber2024}, we described an adaptation algorithm for a mesh organized as a forest of octrees tailored for execution on a single GPU, and a recursive time-stepping scheme for the LBM on this mesh. The algorithm was implemented in 2D and 3D in an open-source C++/CUDA code\footnote{The code repository is hosted on Github: \url{https://github.com/KhodrJ/AGAL}.} called the AGAL framework. We will briefly summarize the organization of the mesh and the solver scheme before presenting the complex geometry components.

    An octree is a tree data structure in which each node is split into exactly eight children. The unique node without a parent is referred to as the root node. A node's level in the tree is the number of parent-child links between it and the root node. Each node corresponds to a single block in the grid composed of $4^D$ cells (Figure \ref{fig:foo}), where $D$ is the number of dimensions. Integer indices, referred to as block IDs, uniquely determine the position of the block and cell data in the solution and metadata arrays. Blocks on a level $L$ are denoted as the grid level $L$. The forest of octrees is enumerated with a set of ID sets $\{\texttt{id\_set}_L\}_{L=0}^{\text{max.}}$, which store the block IDs by grid level.

    The \texttt{Mesh} class implements the forest of octrees structure. The root nodes of all octrees are arranged as a structured grid denoted as the root grid, which can be refined near the domain boundaries before the solver loop takes place, and/or dynamically within the loop. We performed near-wall refinement \cite{Jaber2024} in the lid-driven cavity and flow past a square cylinder benchmark tests by hard-coding the near-wall distance formula according to the known simple geometry (i.e., refining blocks that are a certain distance away from the four/six faces of the square/cubic domain or from the cylinder). Basic bounce-back boundary conditions for the LBM are accurate for this type of geometry but require interpolation or ghost-cell methods when the geometry is not specifically aligned with the cell edges/faces (Figure \ref{fig:lbm_interp}). The procedures outlined in this paper specifically address the issue of meshing (Figure \ref{fig:scope}). Work on the updated LBM scheme is in progress.
    

    \subsection{Representation of the Complex Boundary Geometry}

\begin{figure}[t]
        \centering
        \begin{subfigure}[t]{0.15\textwidth}
            \centering
            \includegraphics[width=1\linewidth]{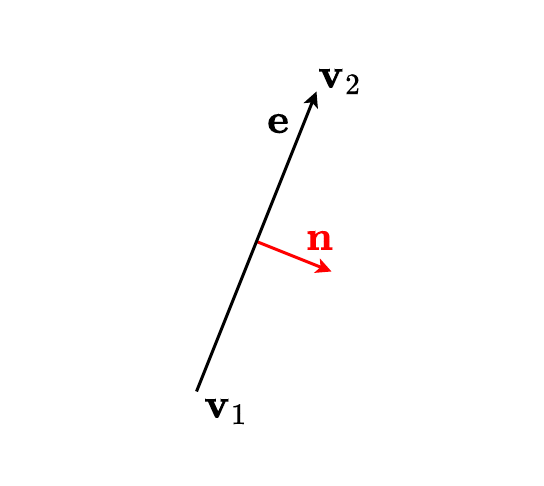}
            \caption{Edge (2D).}
            \label{fig:GE}
        \end{subfigure}
        \begin{subfigure}[t]{0.15\textwidth}
            \centering
            \includegraphics[width=1\linewidth]{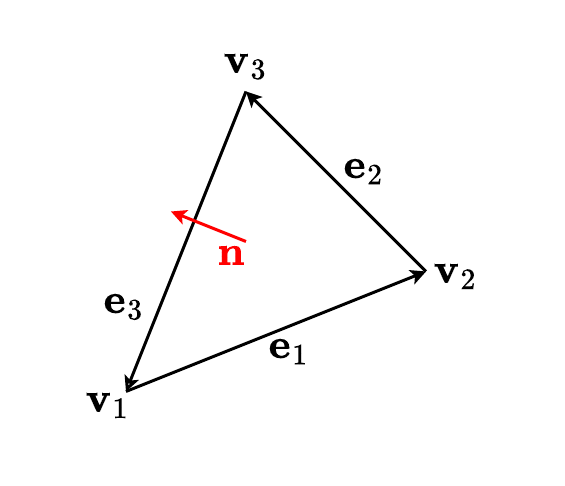}
            \caption{Triangle (3D).}
            \label{fig:GT}
        \end{subfigure}
        \caption{Visualization of the current edge and triangle orientations.}
        \label{fig:TriangleOrientation}
    \end{figure}

\begin{figure}[b]
        \centering
        \includegraphics[width=0.85\linewidth]{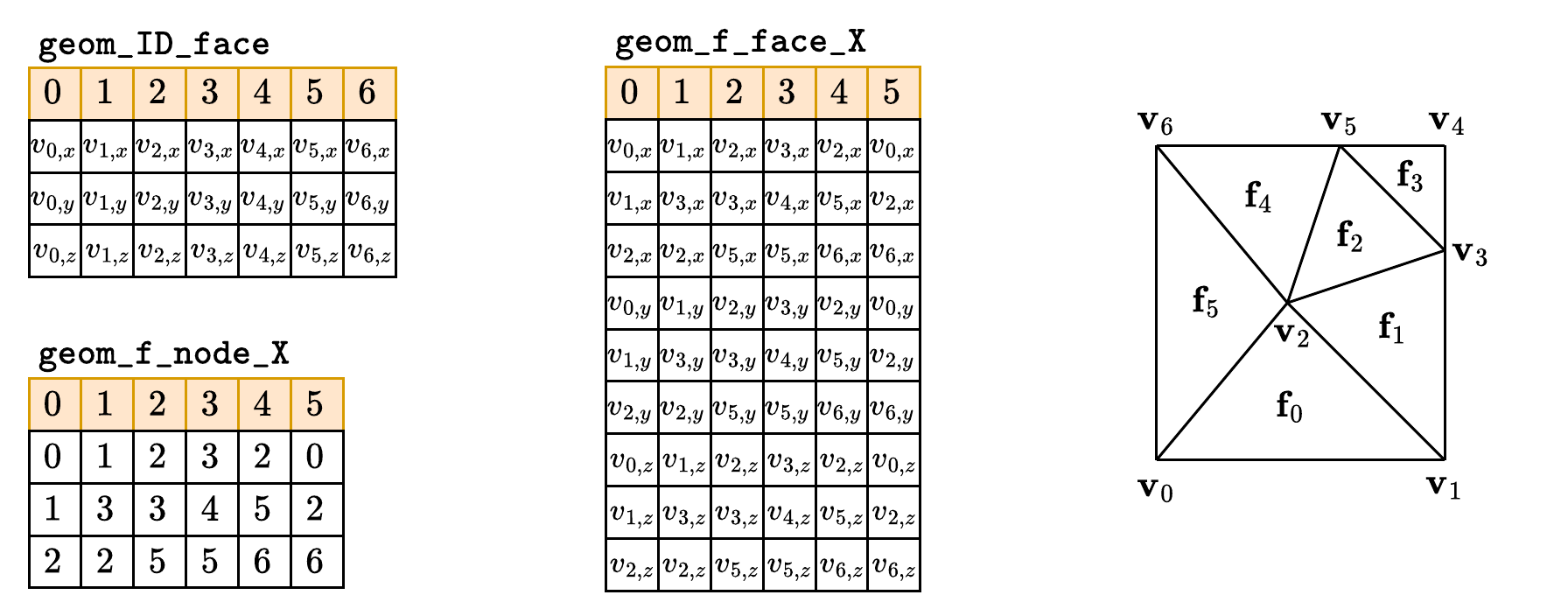}
        \caption{Index ($\texttt{geom\_f\_node\_X}$, $\texttt{geom\_ID\_face}$) and coordinate list ($\texttt{geom\_f\_face\_X}$) representations of a sample triangle mesh. Yellow cells indicate the indices of the elements.}
        \label{fig:data_struct_list}
    \end{figure}

    The geometry is composed of a set of edges/triangles (Figure \ref{fig:TriangleOrientation}) and represented in the form of a set of index lists (i.e., vertex coordinates stored in one array and indices of the edge/triangle vertices in another), or as a single coordinate list for the edges/triangles (which may allow duplicate storage of vertex coordinates). These are illustrated in Figure \ref{fig:data_struct_list}. In this paper, we will refer to both edges and triangles as faces without loss of generality. The routine $\texttt{G\_ImportFromTextFile}$ constructs primitive sub-meshes using user-supplied resolution and location values and appends them to vectors that store the index lists. $\texttt{G\_ConvertIndexListsToCoordsList}$ converts these index lists into a single array of triangle coordinates $\texttt{geom\_f\_face\_X}$ that is organized in a structure of arrays format. This is the more suitable form for the CUDA kernels to be introduced shortly. Alternatively, we provide a routine $\texttt{G\_ImportSTL}$ that reads the vertex coordinates of the faces to directly build the coordinate list. The list is then permanently transferred to the GPU, and it remains accessible by the \texttt{Mesh} object to be passed to the refinement and coarsening and solver routines.

    \subsection{Cell-Face Identification Algorithms}

    \subsubsection{Identification of Cells Near the Boundary}

    
    A naive implementation of the face-detection scheme entails a traversal of the blocks in the primary computational grid, where each block checks if it is adjacent to at least one edge/triangle in the whole geometry. The implementation of e.g., interpolated boundary conditions, requires that cells are able to identify links with the appropriate faces of the domain boundary, compute the distance along a vector defined by the quadrature abscissae in discretizing the Boltzmann equation in particle-velocity space (i.e., a direction in the velocity set), and finally computing the interpolated density distribution function.
    
    Blocks in the grid are traversed as per the so-called primary mode of access as defined in \cite{Jaber2024}. Each grid level is traversed using individual ID sets (i.e., one level at a time), and CUDA threads first read a fixed number of block IDs. Then, in a for-loop, the threads are assigned to each cell in the cell-block and its cells are processed simultaneously. The thread-block size is equal to the cell-block size to ensure that all threads participate in updating the cell-block data.
    
    For near-wall refinement and cell-face link identification, each cell loops over the known number of faces and, for each face, the vertex coordinates are loaded from \texttt{geom\_f\_face\_X}. A face detection algorithm commences whereby the cell checks if its center is a specified distance away from the face. If so, the cell’s block is marked for refinement. The cell can potentially store the faces that is identifies in a separate list for future use in the solver routines as well. The refinement and coarsening procedure (\texttt{M\_RefineAndCoarsenBlocks}, detailed in \cite{Jaber2024}) then subdivides the marked blocks.
    

    \begin{algorithm}[b]
        \small
        \caption{Check if point is in range of a triangle}
        \label{alg:blocktri_range}
        \begin{algorithmic}[1] 
            \Procedure{CheckNearTriangle}{$\textbf{x}_p, R$} 
                \ForAll{$k \in \{1,2,3\}$}
                    \If{$\left(\norm{\textbf{x}_p - \textbf{v}_k}^2 \leq R\right)$} \Comment{Sphere}
                        \State $\textbf{return true}$
                    \EndIf
                    \State $\displaystyle d^2 \gets \frac{\norm{(\textbf{v}_{k+1} - \textbf{v}_k) \times (\textbf{v}_k - \textbf{x}_p)}^2}{\norm{\textbf{v}_{k+1}-\textbf{v}_k}^2}$
                    \State $d_{e,1} \gets -(\textbf{v}_k - \textbf{x}_p)\cdot\textbf{e}_k$
                    \State $d_{e,2} \gets (\textbf{v}_{k+1}-\textbf{x}_p)\cdot\textbf{e}_k$
                    \If{$\left(d^2 \leq R \textbf{ and } d_{e,1}, d_{e,2} \geq 0  \right)$} \Comment{Cylinders}
                        \State \textbf{return true}
                    \EndIf
                \EndFor
                \State $C_1 \gets \bigwedge\limits_{k=1}^3 \left(-(\textbf{v}_k - \textbf{x}_p)\cdot(\textbf{e}_k\times \textbf{n})\geq 0\right)$
                \State $C_2 \gets \left(-(\textbf{v}_1-R\textbf{n}-\textbf{x}_p)\cdot\textbf{n}\right) \wedge \left((\textbf{v}_1 + R\textbf{n}-\textbf{x}_p)\cdot \textbf{n}\right)$
                \If{$C_1 \textbf{ and } C_2$} \Comment{Prism}
                    \State \textbf{return true}
                \EndIf
            \EndProcedure
        \end{algorithmic}
    \end{algorithm}

    \subsubsection{Face Detection}

\begin{figure}[t]
        \centering
        \begin{subfigure}[t]{0.225\textwidth}
            \centering
            \includegraphics[width=1\linewidth]{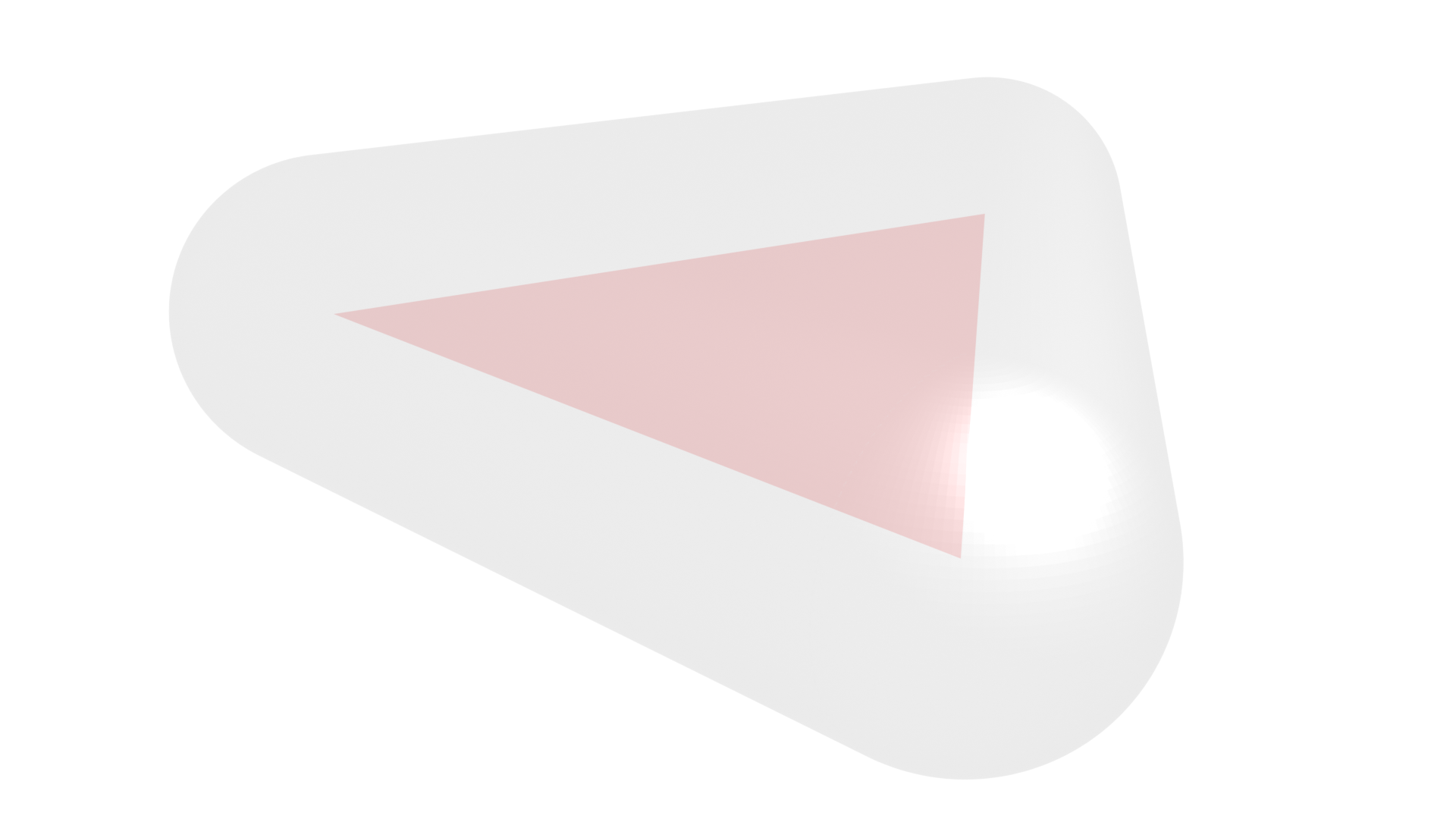}
            \caption{Near-wall region.}
            \label{fig:CC}
        \end{subfigure}
        \begin{subfigure}[t]{0.225\textwidth}
            \centering
            \includegraphics[width=1\linewidth]{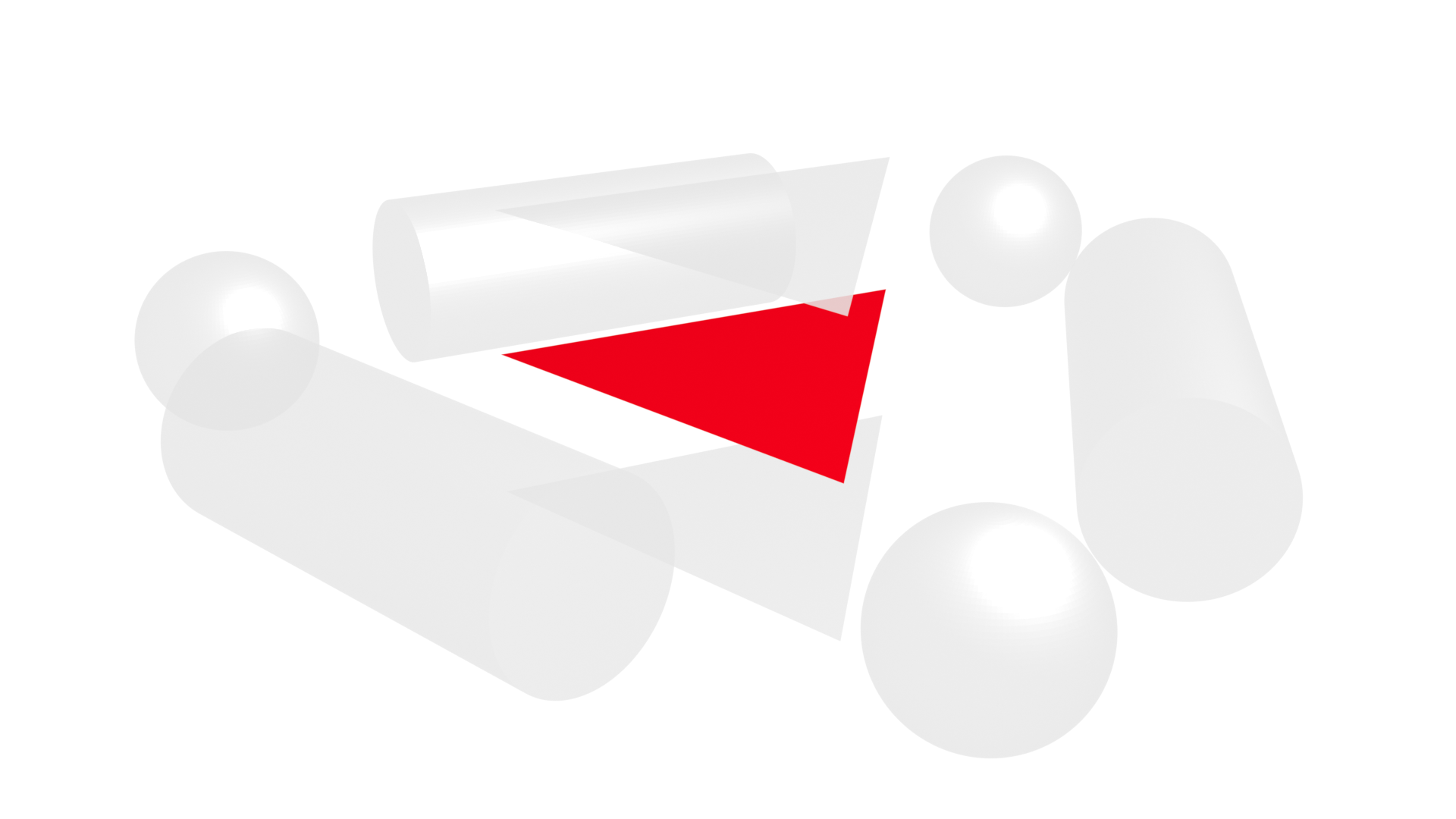}
            \caption{Subdomains to be checked.}
            \label{fig:DD}
        \end{subfigure}
        \caption{Visualization of the near-wall region around a single triangle oriented in 3D.}
        \label{fig:TriangleRange}
    \end{figure}

    A point is considered adjacent to a face if the distance between the two is less than the user-specified near-wall distance $d_{\text{spec.}}$. For a triangle, this requires that the distances from the point to both the planar region and the individual edge segments are less than $d_{\text{spec.}}$. The point may be collinear to the plane (e.g., at a corner) but near one of the edge segments, in which case it is also considered adjacent. This is equivalent to checking that the point is in the region defined by three circles centered on the vertices, three cylinders oriented along the edges, and a triangular prism oriented along the plane (Figure \ref{fig:TriangleRange}, Algorithm \ref{alg:blocktri_range}). The radius of the sphere/cylinder and height of the prism are both equal to $d_{\text{spec.}}$. In 2D, this degenerates to a check on two circles centered on the vertices and a rectangular region aligned length-wise with the edge. We use the point-line distance formula \cite{PointLine} when testing against the cylinders, and the half-space test when testing against the triangular prism.


    \subsubsection{Spatial Binning Algorithm}

\begin{figure}[b]
        \centering
        \includegraphics[width=0.6\linewidth, trim={0 0 2.5cm 2.5cm}]{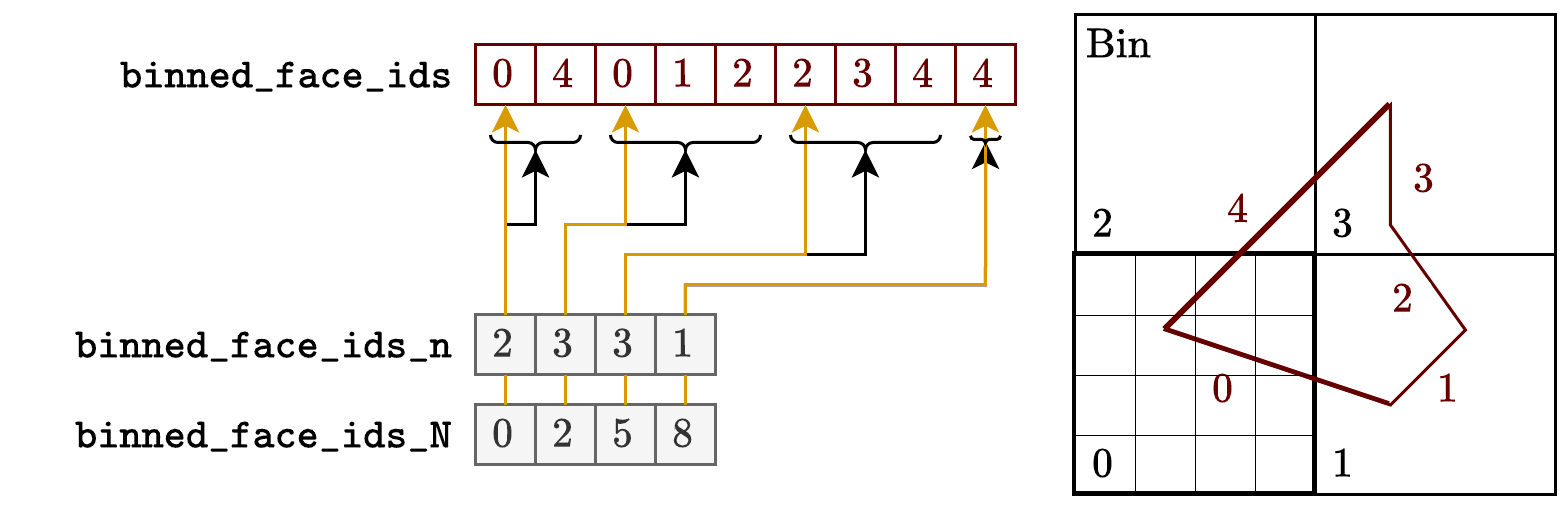}
        \caption{Illustration of the output of the binning procedure.}
        \label{fig:binning_output}
    \end{figure}

\begin{figure}[t]
        \centering
        \begin{subfigure}[t]{0.175\textwidth}
            \centering
            \includegraphics[width=1\linewidth]{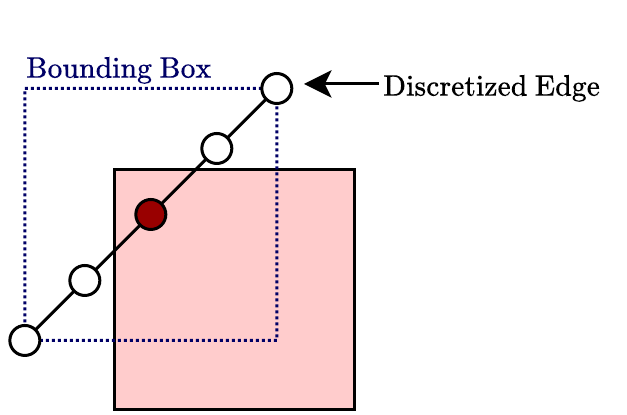}
            \caption{Discretized edge.}
            \label{fig:discrete_edge}
        \end{subfigure}
        \begin{subfigure}[t]{0.175\textwidth}
            \centering
            \includegraphics[width=1\linewidth]{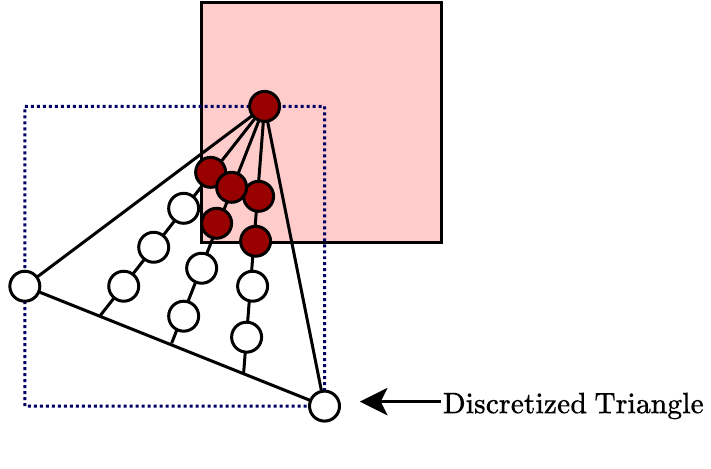}
            \caption{Discretized triangle.}
            \label{fig:discrete_tri}
        \end{subfigure}
        \caption{Discretization of edges and triangles during spatial binning. Discretization of a triangle begins with one edge, followed by the segments formed from the discrete points on the edge and the vertex opposite to it.}
        \label{fig:discrete_elements}
    \end{figure}

    The naive implementation is highly inefficient: accessing all faces from each cell incurs a large global memory access cost, and this is unnecessary since only relatively few cells in the whole domain are in the vicinity of the boundary. We employ a spatial binning approach to reduce the search-loop size (i.e., the set of all faces that need to be searched). The initial computational grid, which is rectangular/prismatic in shape, can be divided into a set of regularly-sized bins. Faces are assigned to bins if they intersect with them or are enclosed entirely by them. While it is possible to check for intersection by using line-plane tests (point-line tests in 2D), we choose to discretize the faces with a number of nodes that (Figure \ref{fig:discrete_elements}) scale with the size of the face, and to check if at least one point lies in the bin. We use three arrays to represent the binning: $\texttt{binned\_face\_ids}$, which stores the face IDs of each bin in order by group, $\texttt{binned\_face\_ids\_n}$, the number of faces in each bin, and $\texttt{binned\_face\_ids\_N}$, the indices in $\texttt{binned\_face\_ids}$ where the first face of each bin is located (Figure \ref{fig:binning_output}).

    The implementation is characterized by bin density $B$, the number of bins to consider along each Cartesian axis (for a total of $N_{\text{bins}} = B^D$, and bin fraction $B_f$, the number of bins to update at a time in the CUDA kernel. The bin fraction determines the amount of memory to allocate in GPU memory. A larger value means that the workload is divided a greater number of times so that fewer bins are updated at a time (which requires a smaller allocation). The bin length is obtained by normalizing the domain size by the bin density. Since faces may be duplicated when crossing bins, a larger amount of memory is allocated to ensure that there is enough space at the beginning. We chose 10x the number of faces, though experiments should that the overlap is relatively minimal. A bin indicator array is also defined for the faces in a structure of arrays format such that each face can indicate occupancy in up to $B_n$= $N_{\text{bins}}/B_f$ bins.
    
    Looping over $B_f$, an amount $B_n$ are updated with the \texttt{Cu\_FillBins} kernel. The face coordinates list is traversed with CUDA threads assigned to each face, and each face loops over $B_n$ bins. If a face is detected as occupying a bin, it is assigned by modifying the appropriate indicator with the face ID. After the kernel execution is completed, a stream compaction is performed where the face IDs in the indicator array are copied into $\texttt{binned\_face\_ids}$ to ensure that they are contiguous in memory. We use Thrust's \cite{Thrust} $\texttt{count\_if}$ and $\texttt{copy\_if}$ to count the number of IDs to be copied and then to perform the copy, respectively.

\begin{figure}[t]
        \centering
        \includegraphics[width=0.85\linewidth]{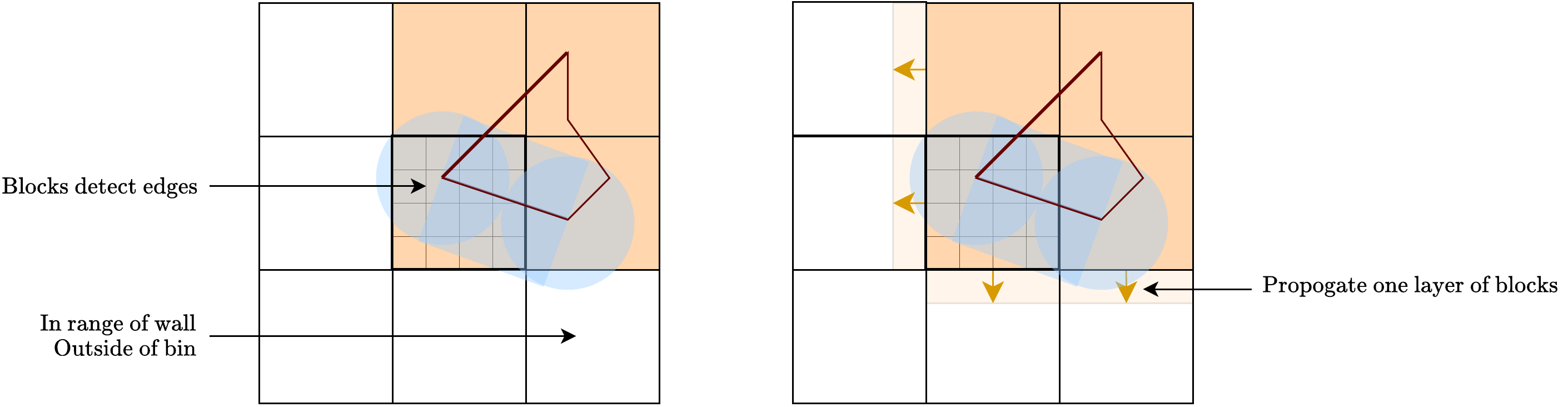}
        \caption{Illustration of refinement mark propagation for binning with a large near-wall refinement distance criterion.}
        \label{fig:propagation}
    \end{figure}

    When the number of bins is increased, cells within a large specified near-wall distance may fail to detect faces. To meet the requirement, we introduce a propagation step (Figure \ref{fig:propagation}) where blocks marked for refinement check their neighboring blocks and mark them for refinement as well until a total distance equal to $d_{\text{spec.}}$ has been covered (i.e., a total of $N_{\text{prop.}} = 1+\lfloor d_{\text{spec.}} / \Delta x_b \rfloor$ are performed, where $\Delta x_b$ is the cell-block length). The propagation is done in two steps to avoid a race condition: the refinement IDs of the nearby blocks are first switched to an intermediate value, and then the intermediate values are replaced with a mark for refinement in a second traversal. The first step follows the so-called secondary mode of memory access as defined in \cite{Jaber2024}.

\section{RESULTS AND DISCUSSION}

\begin{figure}[t]
    \centering
    \begin{subfigure}[t]{0.18\textwidth}
        \centering
        \includegraphics[width=1\linewidth, trim={0 1cm 0 1cm}]{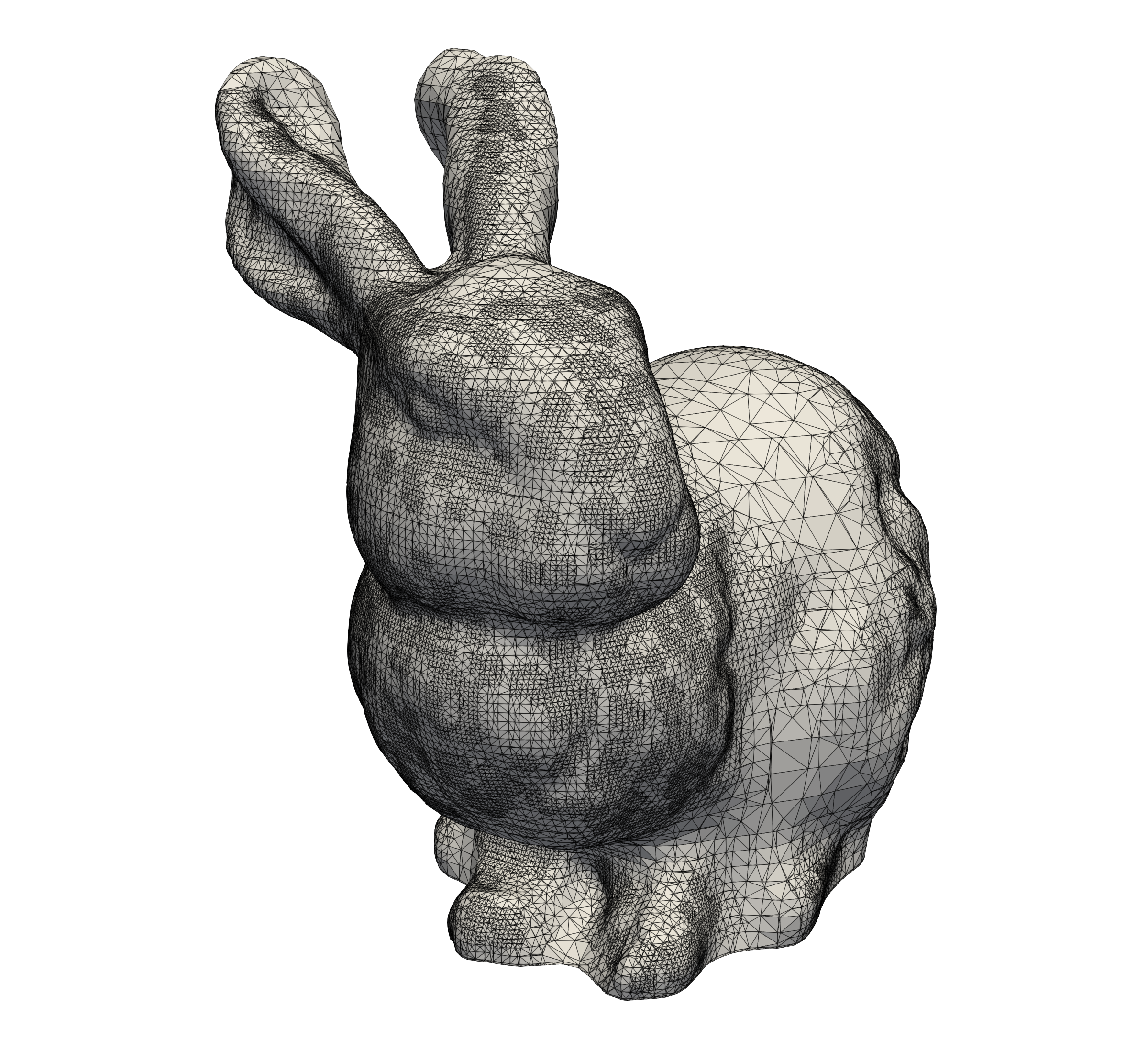}
        \caption{The Stanford bunny.}
        \label{fig:bunny}
    \end{subfigure}
    \begin{subfigure}[t]{0.18\textwidth}
        \centering
        \includegraphics[width=1\linewidth, trim={0 1cm 0 1cm}]{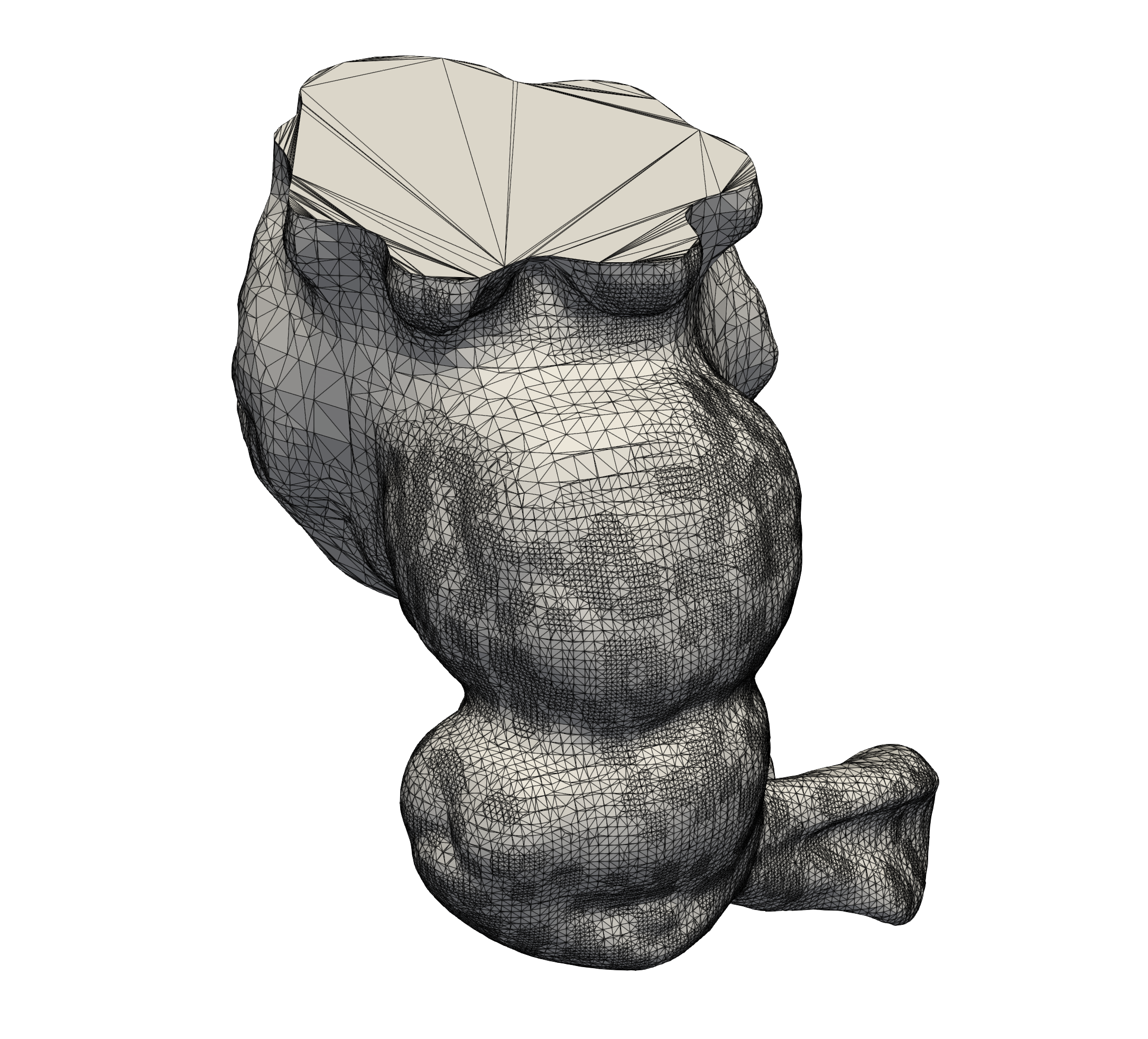}
        \caption{Upside-down view.}
        \label{fig:bunny_ud}
    \end{subfigure}
    \caption{STL model of the Stanford bunny with 112,402 faces.}
    \label{fig:BunnyOriented}
\end{figure}
We record the execution times of the implemented routines for two test geometries to assess the performance of the implementation: a 2D circle and a 3D Stanford bunny \cite{Bunny, BunnyII}\footnote{The Stanford bunny is a well-known benchmark problem in the field of computer graphics. The original model \cite{Bunny} is in PLY format; we used a binary STL model obtained from Wikimedia Commons \cite{BunnyII} and transformed it to ASCII STL with the Blender software.} (Figure \ref{fig:BunnyOriented}). Serial CPU implementations of the face-detection scheme complement the CUDA kernels, which we use to obtain an estimate of the speedup provided. We verify that the number of blocks marked for refinement with the CPU and GPU codes are identical given the same input to validate the former. The code is compiled in single-precision and the tests are executed on a single NVIDIA GeForce 970M with 3GB of VRAM.

\begin{figure}[t]
    \centering
    \begin{subfigure}[t]{0.241\textwidth}
        \centering
        \includegraphics[width=1\linewidth, trim={0 1.2cm 0 1.2cm}]{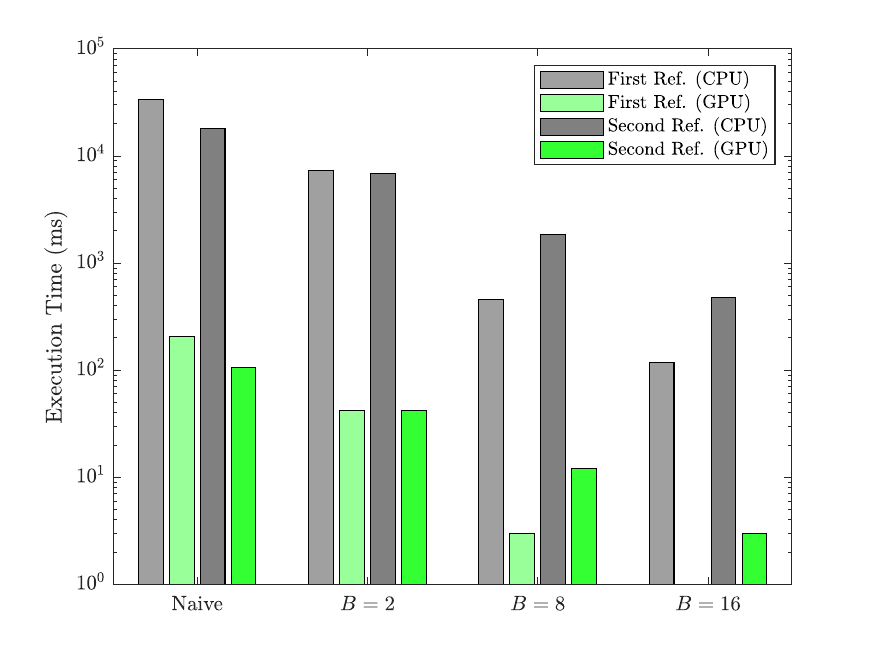}
        \caption{2D.}
        \label{fig:cpuvsgpu_2d}
    \end{subfigure}
    \begin{subfigure}[t]{0.241\textwidth}
        \centering
        \includegraphics[width=1\linewidth, trim={0 1.2cm 0 1.2cm}]{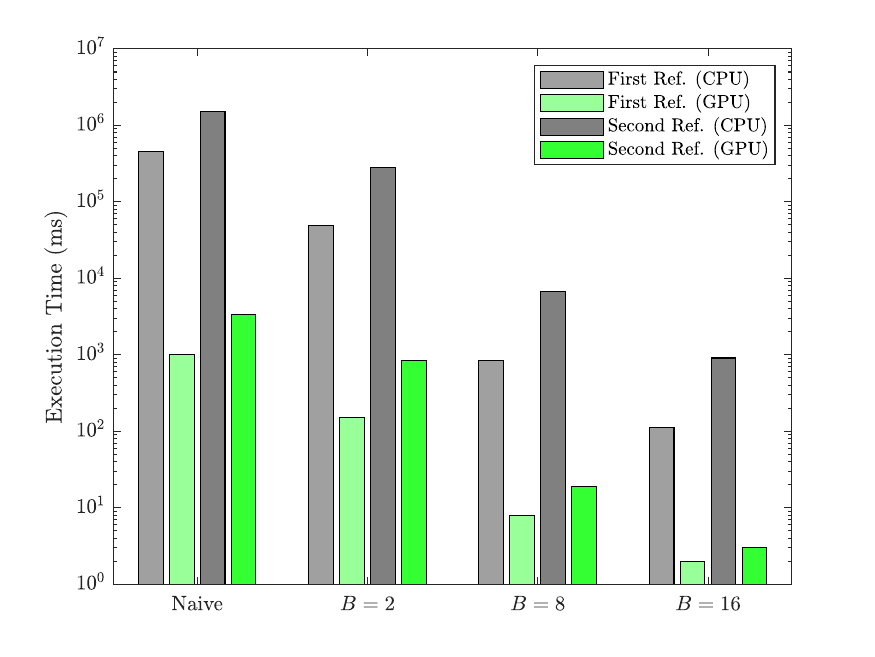}
        \caption{3D.}
        \label{fig:cpuvsgpu_3d}
    \end{subfigure}
    \caption{Distribution of execution times for the near-wall refinement routine on the CPU and GPU.}
    \label{fig:cpuvsgpu}
\end{figure}

Two individual speedup tests are performed for the circle and the bunny. The circle is instantiated with 12,800 edges in the middle of a computational grid with resolution $256^2$ and size $[0,1]^2$, and the near-wall distance criterion is set equal to $d_{\text{spec.}}=0.1$. The grid is then refined twice, for a total of three levels in the grid hierarchy. The bunny possesses 112,402 faces and is placed inside a grid with resolution $64^3$ and size $[0,1]^3$ with $d_{\text{spec.}}=0.05$. Two successive refinements are also used. We consider bin densities $B \in \{1,2,8,16\}$, with $B=1$ indicating the use of the naive scheme. Figures \ref{fig:cpuvsgpu_2d} and \ref{fig:cpuvsgpu_3d} illustrate the difference in order of magnitude in the execution times for the face-detection portion of the refinement.

\begin{figure}[b]
    \centering
    \begin{subfigure}{0.241\textwidth}
        \centering
        \includegraphics[width=1\linewidth, trim={0 .5cm 0 .5cm}]{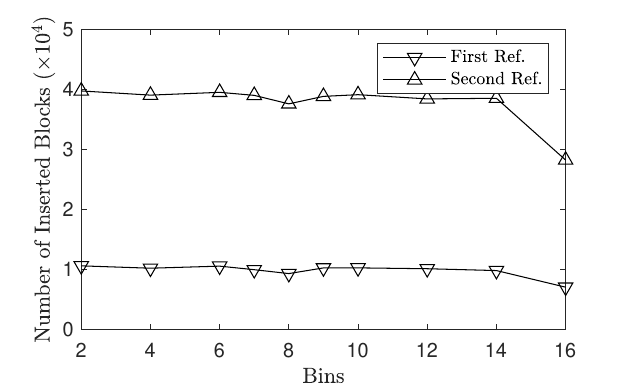}
    \end{subfigure}
    \begin{subfigure}{0.241\textwidth}
        \centering
        \includegraphics[width=1\linewidth, trim={0 .5cm 0 .5cm}]{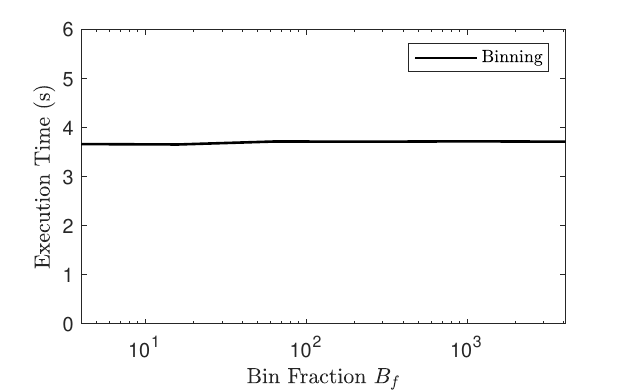}
    \end{subfigure}
    \caption{Left: Inserted blocks vs. bin density. Right: Binning execution time vs. bin fraction.}
    \label{fig:discrete_binning}
\end{figure}

\begin{figure}
    \centering
    \includegraphics[width=0.6\linewidth, trim={0 .35cm 0 .35cm}]{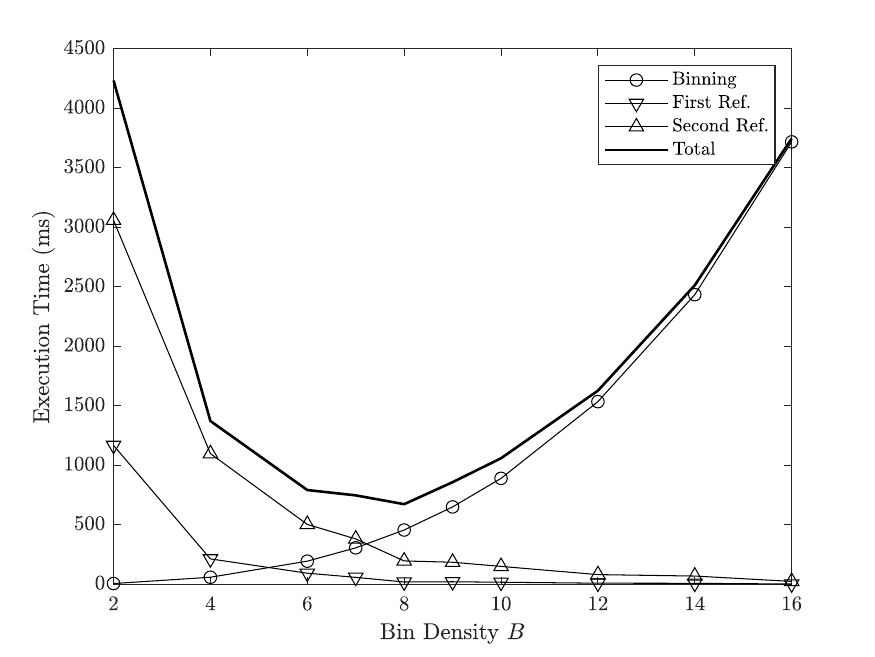}
    \caption{Execution times of Test 2, plotted against bin density.}
    \label{fig:timevsbins}
\end{figure}

\begin{figure}[b]
    \newcommand{\TR}{8}
    \centering
    \begin{subfigure}[t]{0.15\textwidth}
        \centering
        \includegraphics[width=1\linewidth,trim={0 \TR cm 0 \TR cm}]{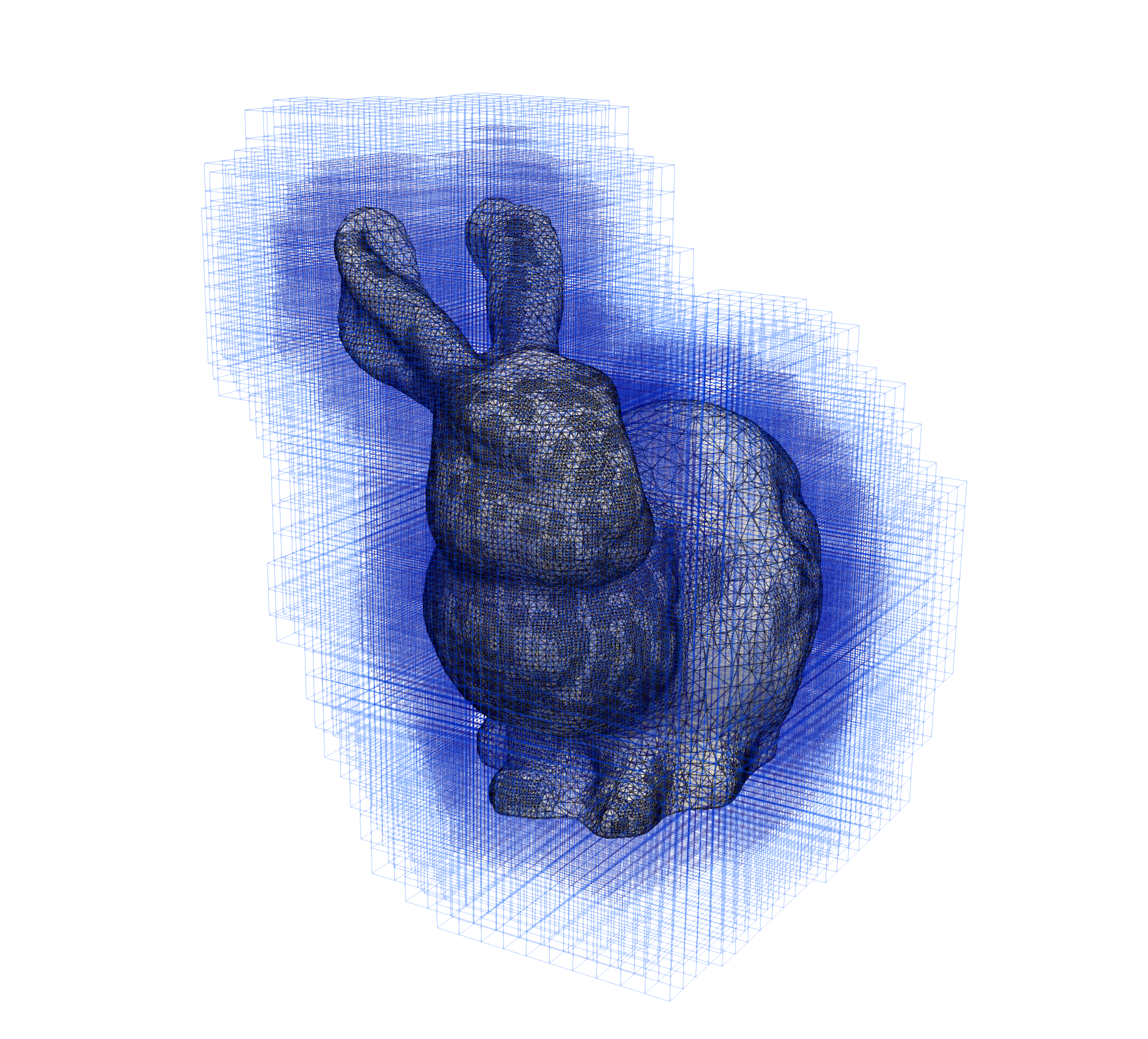}
        \caption{$B = 2$.}
        \label{fig:bunny}
    \end{subfigure}
    \begin{subfigure}[t]{0.15\textwidth}
        \centering
        \includegraphics[width=1\linewidth,trim={0 \TR cm 0 \TR cm}]{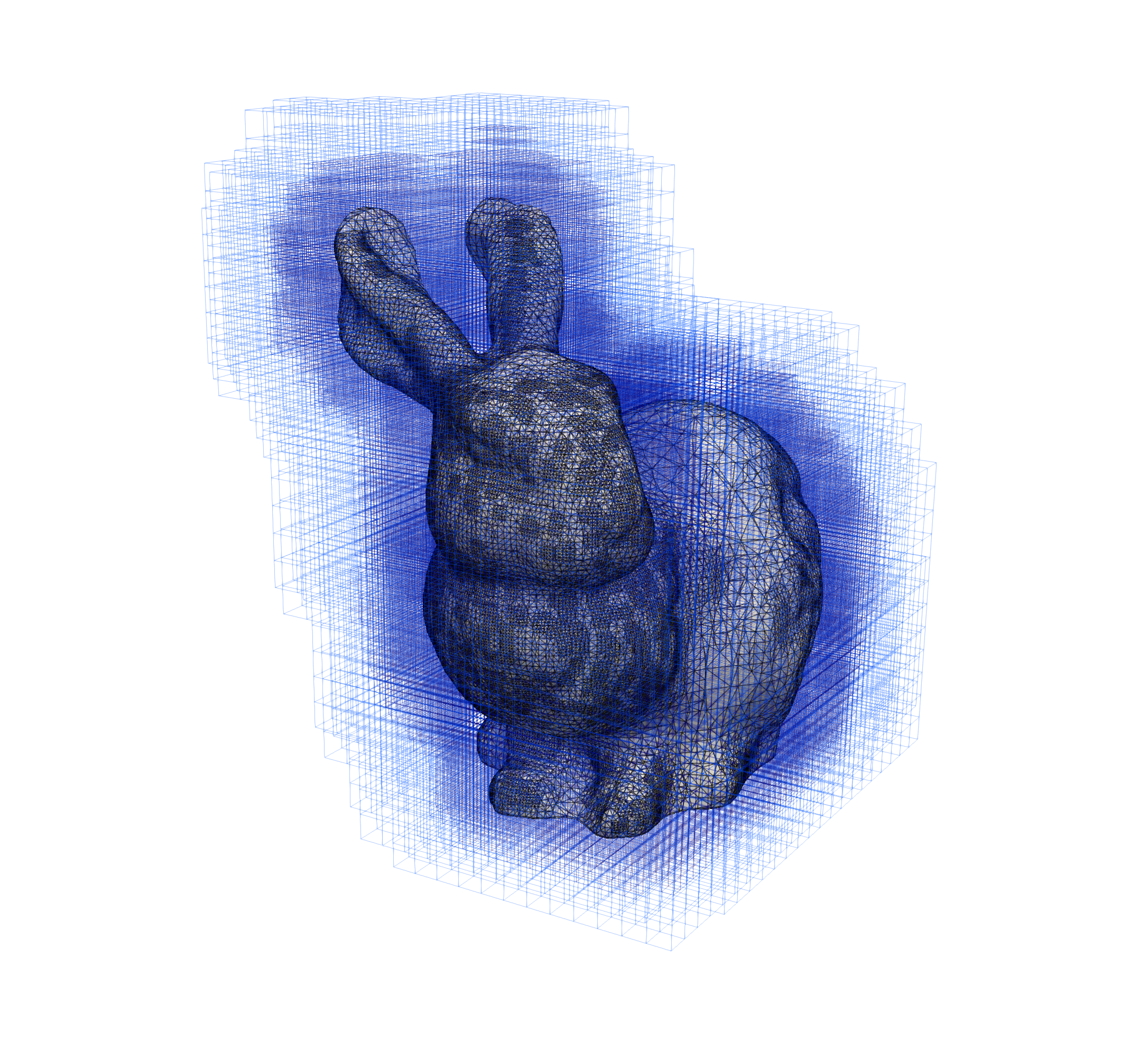}
        \caption{$B = 8$.}
        \label{fig:bunny_ud}
    \end{subfigure}
    \begin{subfigure}[t]{0.15\textwidth}
        \centering
        \includegraphics[width=1\linewidth,trim={0 \TR cm 0 \TR cm}]{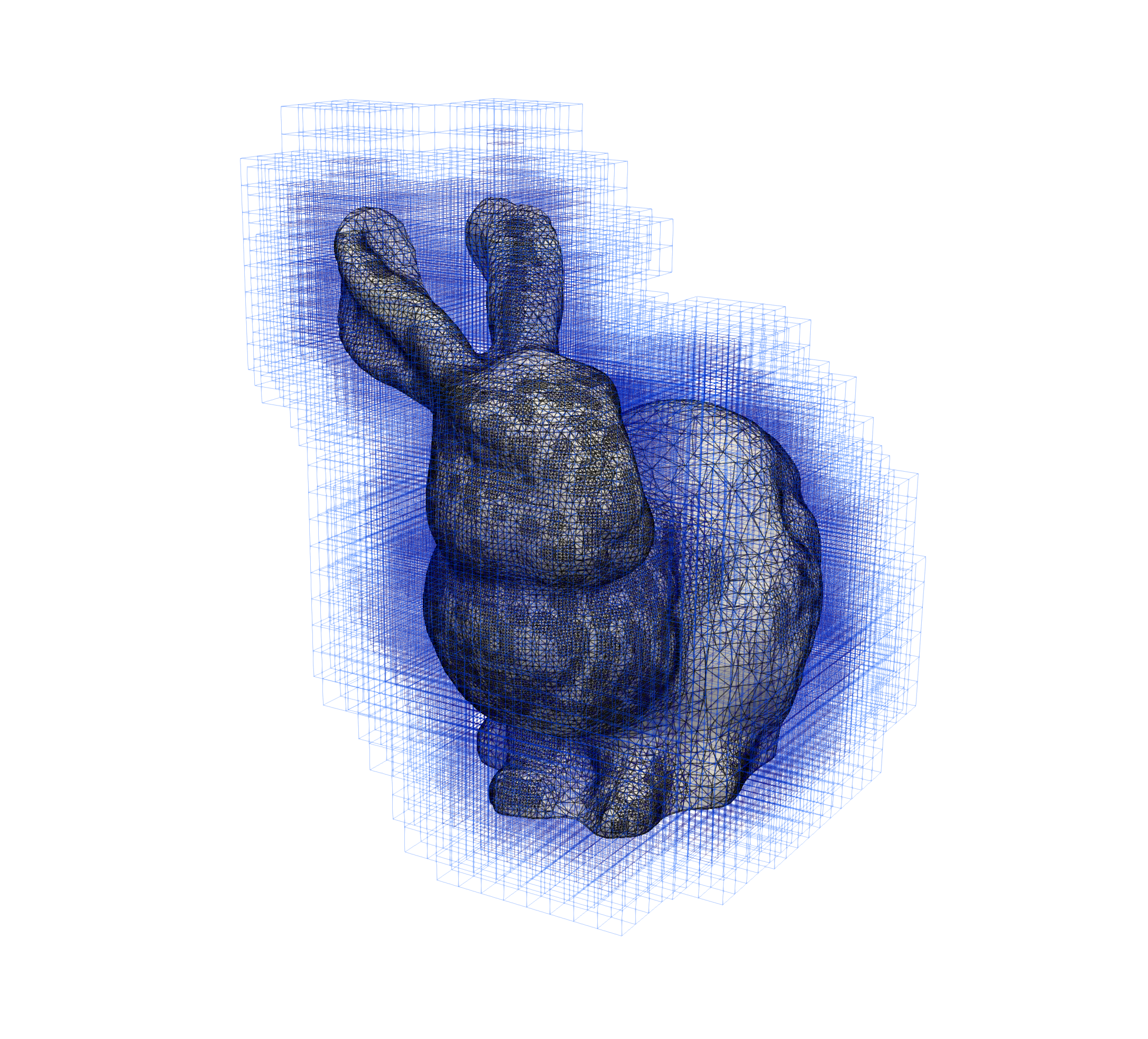}
        \caption{$B = 16$.}
        \label{fig:bunny_ud}
    \end{subfigure}
    \caption{Cell-blocks near the bunny after successive near-wall refinement in the execution time vs. bin density test. As bin density increases, more propagation is needed to satisfy the near-wall refinement distance criterion.}
    \label{fig:BunnyNearWall}
\end{figure}

To test the efficiency of the binning algorithm, we use the same setup for the 3D CPU-GPU comparison, and we vary the bin density $B =\in \{1,2,4,6,7,8,9,10,12,14,16\}$. The variation in execution times for bin setup, face-detection, and the total time is plotted against $B$ in Figure \ref{fig:timevsbins}. We also test the effects of $B$ on the final number of blocks marked for refinement, and the effects of $B_f$ on the execution time of the binning routine (Figure \ref{fig:discrete_binning}). The near-wall refinement results for the Stanford bunny are displayed in Figure \ref{fig:BunnyNearWall}.

The GPU code is consistently two orders of magnitude faster than the CPU counterpart for face-detection operations on both grid levels. As $B$ increases, the execution times decrease for both refinements on both devices, reaching $\mathcal{O}(1)$ \text{ms} at $B=16$ on the GPU and $\mathcal{O}(100)-\mathcal{O}(1000)$ on the CPU. The speedup is at a maximum for the naive algorithm, at around 450x in 3D and 160x in 2D. The speedup is less pronounced in 2D, likely due to the relatively smaller workloads. When the execution time for setting up the bins is accounted for, we find that the total time needed for the near-wall refinement is at a minimum when $B=8$ with a value of approximately $1$ second. More time is spent checking adjacency to faces for smaller $B$, while for larger values the bin setup time quickly overtakes the nearly negligible face-detection time. The total times for $B=2$ and $B=16$ are about the same, equal to $\sim 3.5 - 4$ \text{s}. We found that $B_f$ did not impact the performance of the bin setup procedure, however, larger values were required when $B$ was increased in order to not run out of memory.

The binning procedure is a relatively expensive step as $B$ increases. This may be attributed to the face discretization procedure that is used to assign faces to the bins. The discretization parameter is a quick way of assigning faces to bins, but experiments show that it can impact the total computational cost if the number is too large. There are other ways of determining if a face crosses a bin (e.g., checking that the edge(s) of the face intersect with the faces of the bin, or if the face is totally enclosed by the bin). We have also only considered static refinement of the grid prior to simulation. It is likely that the additional cost of bin setup with a greater $B$ offsets the cost of increasing the search-loop size in the case of a moving geometry (such as with immersed boundaries).

The naive implementation of near-wall refinement makes it easier to classify a node as a solid, boundary, or fluid node since all faces in the domain are available. Ray-casting can be used to determine the node type, and the near-wall distance can be enforced exactly. The binning procedure makes the process much more local, which means different approaches to node classification are needed (such as checking the position of the node relative to the triangle orientation). This, along with the possible optimizations to the binning procedure, will be explored in future work.

\section*{Conclusion}


We present an algorithm for incorporating a complex geometry in the form of an edge/triangle mesh within an adaptive Cartesian grid that is organized hierarchically as a forest of octrees. The C++/CUDA implementation of the algorithm extends the AGAL framework for GPU-native adaptive mesh refinement in the form of a new \texttt{Geometry} module as a first step toward simulating weakly-compressible fluid flow with the Lattice Boltzmann Method. We define a metric for indicating adjacency of grid cells to the wall, and describe a set of CUDA kernels that enables the cells to identify the indices of the faces (which is a necessary step for establishing links with the boundary domain for application of boundary conditions) and to mark blocks in a specified vicinity of the wall for refinement. We present a naive approach where all cells in the grid check their relative position to all faces in the geometry, and a more efficient approach based on spatial binning which greatly reduces the search-loop size.
 
A comparison of the execution times of the bin setup and face-detection routines assesses the performance of the implementation. The algorithms are implemented both on the CPU and the GPU to determine the relative speedup provided by the latter. A 2D circle and 3D Stanford bunny serve as test geometries. The former is imported by text file with uniform edge sizes, while the latter possesses a large, fixed number of triangles of varying size that make it a suitable candidate for validation of the binning algorithm. The GPU code executes two orders of magnitude faster than the CPU code, even when the naive face-detection approach is used. It takes $\mathcal{O}(1)-\mathcal{O}(100)$ ms to execute the face-detection procedure on the GPU with spatial binning (with larger numbers of bins corresponding to shorter times), and $\mathcal{O}(1000)$ ms with the naive approach. Bin setup generally becomes more expensive when the number of bins increases. Total time was minimized when $8^3$ bins were used, producing a total of $\sim 1000$ ms to set up the bins.

We are currently preparing a Lattice Boltzmann solver that is capable of utilizing the data structures presented in this paper to efficiently simulate weakly-compressible fluid flow in complex geometries. Flows past a circle and a sphere will serve as benchmark tests for validation.

\section*{Acknowledgment}
The authors gratefully acknowledge support from the Natural Sciences and Engineering Research Council of Canada, Canadian Microelectronics Corporation and the Digital Research Alliance of Canada.





\end{document}